# Volumetric Breast-Density Measurement Using Spectral Photon-Counting Tomosynthesis: First Clinical Results

Erik Fredenberg[1,*], Karl Berggren[1], Matthias Bartels[2], and Klaus Erhard[2]

[1] Philips Health Systems, Mammography Solutions, Smidesvägen 5, 17122 Solna, Sweden
[2] Philips Research, Röntgenstrasse 24-26, 22395 Hamburg, Germany
* erik.fredenberg@philips.com

**Abstract.** Measurements of breast density have the potential to improve the efficiency and reduce the cost of screening mammography through personalized screening. Breast density has traditionally been evaluated from the dense area in a mammogram, but volumetric assessment methods, which measure the volumetric fraction of fibro-glandular tissue in the breast, are potentially more consistent and physically sound. The purpose of the present study is to evaluate a method for measuring the volumetric breast density using photon-counting spectral tomosynthesis. The performance of the method was evaluated using phantom measurements and clinical data from a small population ($n = 18$). The precision was determined to 2.4 percentage points (pp) of volumetric breast density. Strong correlations were observed between contralateral ($R^2 = 0.95$) and ipsilateral ($R^2 = 0.96$) breast-density measurements. The measured breast density was anti-correlated to breast thickness, as expected, and exhibited a skewed distribution in the range [3.7%, 55%] and with a median of 18%. We conclude that the method yields promising results that are consistent with expectations. The relatively high precision of the method may enable novel applications such as treatment monitoring.

## 1 Introduction

It is well established that breast density is directly correlated to the risk of developing breast cancer [1], and anti-correlated to the diagnostic accuracy of mammography [2]. Measures of breast density can therefore improve risk estimates and may enable personalized breast-cancer screening, which in turn has the potential to increase the sensitivity and minimize the cost of screening programs [3]. Other areas of application for breast-density measures include treatment monitoring and dose estimation.

Traditionally, breast density has been estimated from the areal fraction of the breast that is covered by fibro-glandular tissue in the mammogram, either visually (e.g. BI-RADS scoring) or automatically. Volumetric breast-density assessment methods measure the volumetric fraction of fibro-glandular tissue in the breast and are becoming an established alternative to traditional methods because of a physically more meaningful interpretation and higher consistency [4].

A number of methods have been developed to measure the volumetric breast density [5–7]. Most often, additional information and assumptions are required in addition to the mammogram itself, such as the compression height [5, 6], a reference pixel value [7], and / or a breast model to take any thickness gradient into account [6], which add to the uncertainty of the measurement.

Two current trends in x-ray imaging of the breast are spectral imaging and tomosynthesis. Spectral imaging has been applied to measure breast density in two-dimensional (2D) mammograms [8, 9]. The technique can be expected to be more accurate than non-spectral methods because there is no need for additional assumptions or modelling. Tomosynthesis is three-dimensional (3D) imaging from a limited angular span. Coupled with spectral imaging, tomosynthesis has the potential to improve breast-density measurements in several respects, including improved precision and 3D localization.

Recently, a prototype system for spectral tomosynthesis has been developed by Philips Health Systems (Solna, Sweden), which is based on the same photon-counting technology that has previously been developed for 2D mammography [10]. In this study, we present a spectral breast-density measurement method developed for the photon-counting tomosynthesis system. The method is evaluated using phantom measurements and the first clinical data acquired with the system.

## 2      Materials and Methods

### 2.1    Spectral Photon-Counting Tomosynthesis System

The Philips MicroDose S0 is a prototype spectral photon-counting tomosynthesis system based on the Philips MicroDose SI 2D mammography system (Philips Health Systems, Sweden). The system comprises a tungsten-target x-ray tube with aluminum filtration, a pre-collimator, and an image receptor, all mounted on a rigid scan arm (Figure 1, left). To acquire an image, the scan arm is rotated around a point below the patient support so that the tube-collimator-detector assembly is scanned across the object.

The image receptor consists of 21 photon-counting silicon strip detector lines with corresponding slits in the pre-collimator (Figure 1, right). During the scan, each detector line will view each point in the object from a unique source-detector angle, and readouts from the 21 lines can therefore be used for 3D reconstruction. The width of the detector and the source-detector distance yield a tomographic angle of ~11°.

Photons that interact in the silicon strip detectors are converted to pulses with amplitude proportional to the photon energy [10]. A high-energy threshold sorts detected pulses into two bins according to energy, which enables spectral imaging. A low-energy threshold provides efficient rejection of electronic noise by discriminating against all pulses below a few keV. The multi-slit geometry rejects virtually all scattered radiation [11]. Low levels of electronic noise and scattered radiation improve the efficiency of tomosynthesis in general and of spectral tomosynthesis in particular.

The 3D reconstruction was iterative (ART), but care was taken to minimize non-linear properties of the algorithm. In this first study we will not consider the 3D properties of the tomosynthesis reconstruction, but the reconstructed stack was summed over the depth direction to form a 2D image.

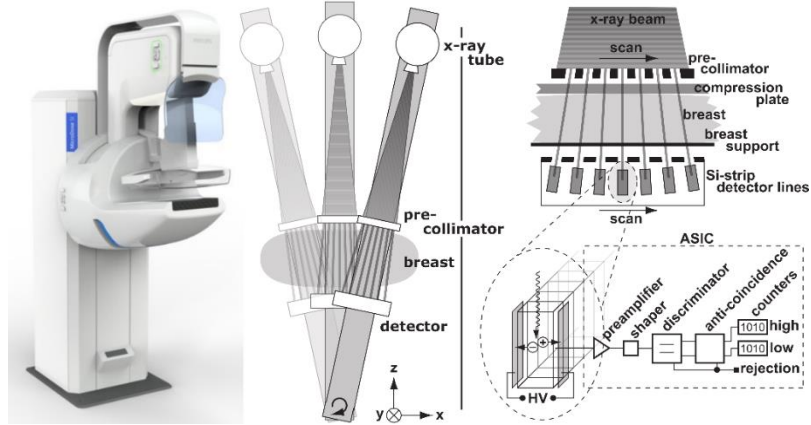

**Fig. 1. Left**: Photograph and schematic of the Philips MicroDose S0 prototype spectral tomosynthesis system. **Right:** The image receptor and electronics.

### 2.2 System Calibration and Spectral Breast-Density Measurement

X-ray attenuation of soft tissue in the mammographic energy range is approximately made up of only two interaction effects, namely photoelectric absorption and scattering processes [12]. Assuming known system properties, constant and known skin thickness, and known linear attenuation coefficients of adipose, glandular and skin tissue, acquisitions over two different energy ranges yield a non-linear system of equations with a unique solution for the thicknesses of adipose and glandular tissue. In practice, this equation is preferably solved with means of calibration because the system and reconstruction properties may be complex and partially unknown. A calibration phantom consisting of combinations of aluminum (Al) and polyethylene (PE) was manufactured for this purpose and imaged at each x-ray energy spectrum (i.e. each kVp) used in clinical practice. From this data set, a look-up table was produced to map image signal for the two energy bins to equivalent amounts of Al and PE. The calibrated Al and PE thicknesses were transferred to equivalent thicknesses of adipose and glandular tissue (attenuation according to Ref. [13]) by using a linear transfer function [14].

The images calibrated to adipose and glandular thicknesses were converted to total breast thickness (sum of adipose and glandular thicknesses) and volumetric breast density (thickness of glandular tissue over total thickness). We refer to these images as thickness and density maps. The average of the density map was taken as the volumetric breast density for the image. The mode (most common value) of the thickness map was used as a measure of breast thickness. We assumed 1.5 mm of skin on each side of the breast. The pectoralis muscle was segmented and excluded from all measurements.

### 2.3 Phantom Measurements and Clinical Measurements

Slabs of tissue-equivalent material (CIRS Inc., Norfolk VA) in 57 different combinations were used to evaluate the precision of the breast-density measurement method. The thicknesses of CIRS adipose and glandular material were converted to equivalent thicknesses of adipose and glandular tissue according to Ref. [13] using a linear transfer function. These values were used as ground truth. Precision was defined as the standard deviation of the differences between measurements and ground truth.

A clinical study of photon-counting spectral tomosynthesis is ongoing at ImageRive, Geneva, Switzerland. Symptomatic patients are examined in two views (CC and MLO) using a MicroDose S0 system. All patients are asked to provide written informed consent prior to the examination. The study has been approved by SwissEthics. Data from the first $n = 18$ patients were included in the present study. One examination was excluded because of technical reasons, which yields a total of 68 measurement points (17 patients × 4 views).

The clinical data was evaluated in terms of: correlation between measured breast thickness and compression height (a strong correlation is expected); correlation between breast density and breast thickness (a weak anti-correlation is expected); the contralateral (left-right) breast density (a strong biological correlation is expected); the ipsilateral (CC-MLO) breast density (a strong correlation is expected).

## 3 Results

Fig. 2 shows the measured volumetric density and thickness as a function of ground truth for the 57 phantom configurations. The precision of the density measurement was 2.4 percentage points (pp). The precision of the thickness measurement was 1.2%.

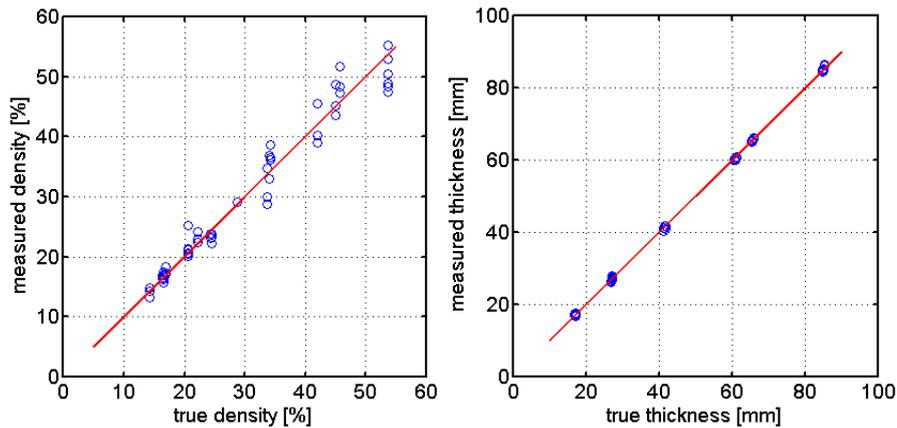

**Fig. 2.** Breast density (left) and thickness (right) as a function of ground truth from 57 phantom configurations. One-to-one lines are included for reference (red solid lines).

The left panel of Fig. 3 shows, as example, the results from the first spectral mammogram that was acquired. The sum of the reconstructed stack, in right CC view, is shown to the left, and the thickness and density maps are shown to the right. This case had a density of 27% and a thickness of 47.2 mm. The densities of the other three views of this patient (expected to be close to identical) were 27% (left CC), 27% (right MLO), and 28% (left MLO). The thicknesses of the other three views (not necessarily identical) were 51.3 mm, 44.1 mm, and 44.4 mm.

The right panel of Fig. 3 shows the measured breast thickness (including the skin) as a function of the compression height reported by the system for all images of all patients (68 images in total). The correlation was strong with a Pearson correlation coefficients of $R^2 = 0.97$. There was a close-to constant offset of $-0.7$ mm.

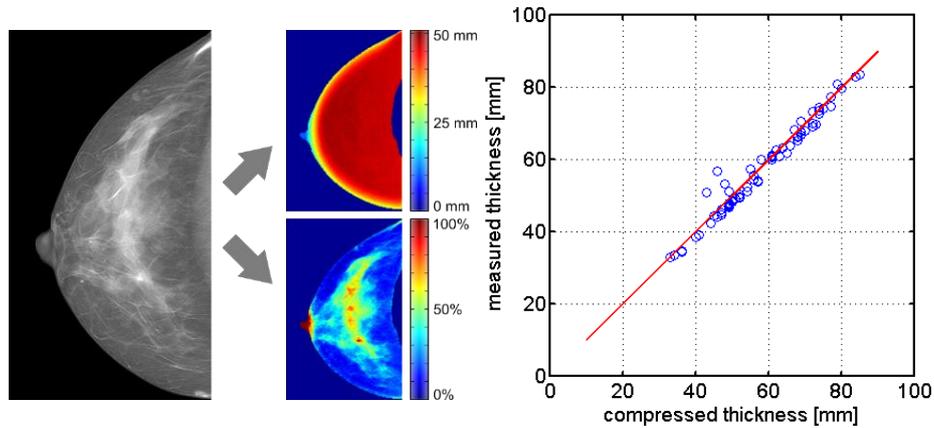

**Fig. 3. Left:** Illustration of the spectral breast-density measurement with the sum of the reconstructed stack, the density map, and the thickness map for a typical case in CC view. **Right:** Compression height as a function of measured breast thickness for 68 images. A one-to-one line (red) is included for reference.

Fig. 4 shows the volumetric breast density, calculated from the 68 spectral mammograms, as a function of measured breast thickness. The data was fitted to a power function of the form [density / %] $= 100 \times a \times$ [thickness / mm]$^b$, where the coefficients were found to be $a = 121$ and $b = -1.58$. Also shown in Fig. 4 are histograms for the distributions of volumetric breast density and breast thickness. The range of densities was [3.7%, 55%], with a mean of 22% and a median of 18%. The range of thicknesses was [33, 83.5] mm, with a mean of 57.8 mm.

Fig. 5 shows contralateral and ipsilateral volumetric breast-density scatter plots. The Pearson correlation coefficients were $R^2 = 0.95$ and $R^2 = 0.96$, respectively. The spread (one standard deviation) around the one-to-one lines were 3.1 pp and 2.8 pp, respectively.

## 4   Discussion

We make the following observations of the results in Sec. 3:

- There was a strong correlation between compression height and measured breast thickness (Fig. 3, right), which is physically sound. The constant offset of less than 1 mm is well within the expected accuracy of the compression paddle.
- As expected, the density was anti-correlated to breast thickness (Fig. 4).

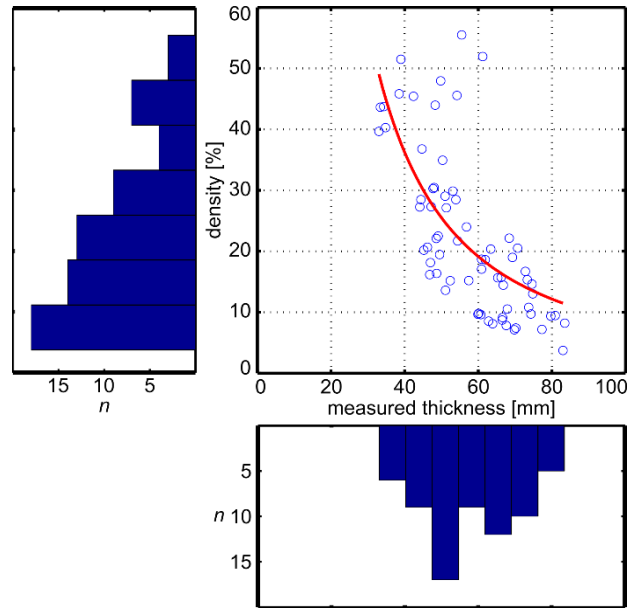

**Fig. 4.** Volumetric breast density from 68 spectral mammograms. The scatter plot shows the dependency on breast thickness, fitted to a power function (red line). The marginal histograms show projections along the axes of the scatter plot, i.e. distributions of densities and thicknesses.

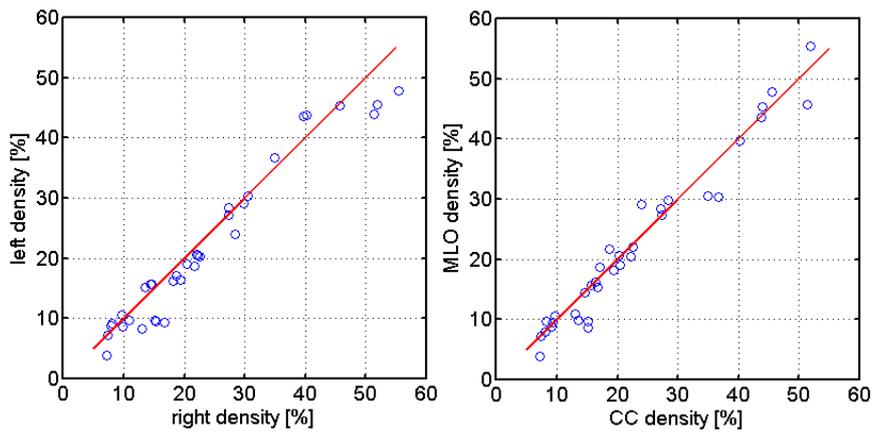

**Fig. 5.** Contralateral (left) and ipsilateral (right) breast density scatter plots. One-to-one lines (red) are included for reference.

- The distribution of breast densities (Fig. 4) was as expected; the distribution was skewed with mean and median close to 20% and a range within [0%, 100%]. The distribution of breast thicknesses (Fig. 4) was also reasonable; the distribution appeared normal (or log-normal) with a mean close to 50 mm. The data are at this point too sparse for any further investigation of distribution statistics.
- The contralateral and ipsilateral correlation measures were relatively strong compared to non-spectral methods (e.g. $R = 0.864 \Rightarrow R^2 = 0.75$ in Ref. [15]).
- The ipsilateral correlation is in close agreement with previous results for 2D spectral mammography ($R = 0.99 \Rightarrow R^2 = 0.98$ in Ref. [16]). The difference is likely covered by differences in population, the lower experience of the clinical staff in positioning for tomosynthesis, and the limited amount of data in the present study.
- The ipsilateral correlation was stronger and the spread was lower compared to the contralateral measures, indicating that the contralateral variation was affected by biological factors in addition to technical factors such as positioning. We therefore expect the ipsilateral measures to be better indicators of system precision.
- In fact, the ipsilateral precision (2.8 pp) was in good agreement with the precision as determined by phantom experiments (2.4 pp). The deviation can be attributed to differences in positioning between CC and MLO views for the clinical images.
- The relatively high precision offered by spectral breast-density measurement may enable applications such as treatment monitoring, where subtle differences in density of typically a few pp indicate response to tamoxifen or raloxifene treatment [17].
- The spectral measurement algorithm presented here is relatively basic and improvements in precision can be expected by taking systematic variations in the spectral response, such as detector and x-ray tube temperature, into account.
- The study population included symptomatic women, which may have affected the results slightly, but we expect this contribution to be non-favorable. For instance, if a lesion were present in one breast, the contralateral correlation would be reduced.
- The present study marks an important and early step in our efforts to combine spectral imaging and tomosynthesis. The study does not cover the advantages that tomosynthesis may offer over 2D mammography for applications such as local breast density measurements, but such applications are part of ongoing research.

## 5    Conclusions

The method for measuring volumetric breast density using spectral photon-counting tomosynthesis was found to yield reasonable results on a small population of women. The precision of the method was determined to 2.4 percentage points (pp) on phantoms, a result that was corroborated by correlation in the clinical data. This relatively high precision (compared to non-spectral methods) may be useful to monitor changes in breast density on an individual basis, for instance in response to treatment.

Research is ongoing to further improve the precision of the method by using a more sophisticated spectral measurement algorithm. Future research will make use of the advantages offered by tomosynthesis for improved local breast-density measurements.